\documentstyle[11pt,AATS,epsf]{article}	

\begin{document}	


\title{Tuning the Clock: Uranium and Thorium Chronometers Applied to
CS~31082-001
}



\author{R. Toenjes, H. Schatz}
\affil{National Superconducting Cyclotron Laboratory and Dept. of 
Physics and Astronomy, Michigan State University}
\author{K.-L. Kratz, B. Pfeiffer}
\affil{Inst. F\"ur Kernchemie, Universit\"at Mainz, Germany}
\author{T.C. Beers}
\affil{Dept. of Physics and Astronomy, Michigan State University}
\author{J. Cowan}
\affil{Dept. of Physics and Astronomy, University of Oklahoma}
\author{V. Hill}
\affil{European Southern Observatory, Garching, Germany}



\begin{abstract}
We obtain age estimates for the progenitor(s) of the extremely metal-poor
([Fe/H $= -2.9$) halo star CS~31082-001, based on the recently reported first
observation of a Uranium abundance in this (or any other) star.  Age estimates
are derived by application of the classical r-process model with updated
nuclear physics inputs. The [U/Th] ratio yields an age of 13$\pm$4 Gyr or
8$\pm$4 Gyr, based on the use of the ETFSI-Q or the new HFBCS-1 nuclear mass
models, respectively.  Implications for Thorium chronometers are discussed.

\end{abstract}

\section{Introduction}

Abundance measurements of long-lived radioactive neutron-capture species found
in r-process-enhanced extremely metal-poor (EMP) stars offer the opportunity to
determine the age of the progenitor object(s) that contributed these elements
into the presently observed stellar material, by comparison with r-process
model predictions (Cowan et al. 1999; Meyer \& Truran 2000). Such age estimates
do not rely on assumptions about Galactic chemical evolution, yet provide
constraints on the nature of r-process nucleosynthesis, and limits on the age
of the Galaxy and the universe.

So far this method has been applied to Thorium observations in several EMP
stars.  For the first time, Uranium now has also been detected, in the EMP halo
star CS~31082-001 ([Fe/H] = --2.9) (Cayrel et al.  2001; also this volume).
Instead of one chronometer -- the previously used [Th/X] abundance ratio (with
X representing a reliably observed stable r-process element such as Eu) --
there are now three principal chronometer pairs available, [Th/X], [U/X], and
[U/Th].  These chronometer pairs can be combined to obtain a more reliable age
estimate (see Goriely \& Clerbaux 1999).  In that way one can also place
another constraint on r-process model parameters by demanding consistency of
the different inferred ages.  R-process model parameters determined in this way
for CS~31082-001 could then be used for more reliable age determinations of
other stars, even those for which no U can be detected.

\section{Calculations}

\begin{figure}[htb]
\plotone{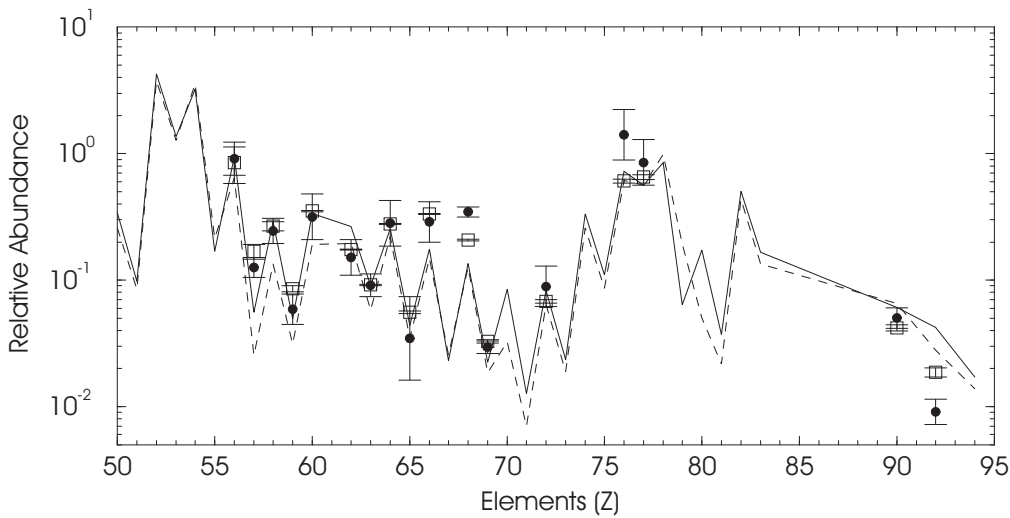}
\caption{Calculated r-process elemental abundances for the ETFSI-Q mass model
(solid) and the HFBCS-1 mass model (dashed) compared to the normalized solar
abundances (open squares) and the observed abundances in  CS~31082-001 (filled
dots). \label{FigAb}}
\end{figure}

We calculated the r-process production of $^{238}$U and $^{232}$Th using the
classical site-independent model (Cowan et al. 1999) assuming
(n,$\gamma$)-($\gamma$,n) equilibrium (waiting-point approximation) and an iron
seed.  R-process abundances are obtained from a superposition of abundance
distributions where weights and timescales obey a power law over the neutron
density.  Assuming that the abundance pattern in CS~31082-001 resembles a solar
r-process pattern (see Fig.~\ref{FigAb}), the four free parameters of the model
are determined by fitting the calculated abundances to the solar system
r-process abundances (Arlandini et al. 1999) at the 124$<$A$<$133 and
189$<$A$<$199 abundance peaks, as well as at Pb.  Because Pb is a product of
$\alpha$-decay chains like Th and U, it provides an important constraint for
the U and Th production estimates (Pb has not yet been detected in
CS~31082-001, Hill et al., this volume).

The necessary nuclear physics input for our calculations are masses, $\beta$-
decay half-lives, branchings for $\beta$-delayed neutron emission, and the
rates of fission processes for very neutron-rich nuclei.  While we used
experimental data where available, the vast majority of the required nuclear
information needs to be predicted by theory. For comparison, calculations were
performed with two nuclear mass models, the frequently used ETFSI-Q (Pearson,
Nayak \& Goriely 1996), and the recently developed HFBCS-1 (Tondeur, Goriely \&
Pearson 2000), which we use here for the first time in a r-process calculation.
The $\beta$-decay data were the same as in Cowan et al. (1999).  It has been
emphasized before that the proper treatment of fission processes is crucial for
calculating the r-process yields of Thorium and Uranium (e.g., Cowan,
Thielemann \& Truran 1991; Goriely \& Clerbaux 1999).  In this work we
re-calculated neutron-induced fission, $\beta$-delayed fission, and spontaneous
fission rates based on the new fission barriers from Mamdouh et al. (1998),
using the methods outlined in Kodoma \& Takahashi (1975).

Ages were determined based on the predicted abundance ratios [U/Th], [U/X], and
[Th/X]. To compensate for the deficiencies in the abundance predictions of
stable r-process elements, and to reduce the impact of observational errors in
the [U/X] and [Th/X] ratios, we also fitted both the observed and the
simulated elemental abundance pattern to the solar one.  Then the ratios
[U*f/U0] and [Th*f/Th0] alone can be used for an age determination.  (U* =
observed stellar abundance, U0 = predicted abundance, f = normalization
factor).

We conducted an extensive investigation of the influence of uncertainties on
the age determination.  (1) $\beta$-decay rates:  To obtain a conservative
estimate of the possible influence of errors in the $\beta$-decay data,
we randomly changed the $\beta$-decay half lives by factors between 0.2 and 5.
Branchings for $\beta$-delayed neutron emission were also varied.  One hundred
calculations were then made to determine the variance of the r-process
abundance predictions ($\Delta_\beta$).  (2) Model uncertainties:  We determined
the range of the four r-process model parameters that still result in a
reasonable fit of the solar abundance pattern. This parameter range leads to an
uncertainty $\Delta_{\rm par}$ in the predicted abundances.  While this error
is small for [U/Th], it turns out to be unacceptably large for the [U/X] and
[Th/X] ages.  However, in this work we constrain the simulation parameters by
requiring best possible consistency with the [U/Th] age.  This constraint leads
to the same small $\Delta_{\rm par}$ for all abundance ratios.  (3)
Observational errors:  These lead to uncertainties in the derived abundance
ratios $\Delta_{\rm exp}$.  (4) Mass models:  We determined the simulation
parameters individually for the two mass models ETFSI-Q and HFBCS-1.  Generally
there is good agreement in the abundance predictions between the two
calculations in the critical 230$<$A$<$260 mass region.  Yet, the pronounced
drop in the calculated abundances in the mass region 229$<$A$<$238 before
$\beta$-decay occurs 5 mass units earlier for the  ETFSI-Q model, resulting in a
somewhat lower Th abundance and a higher age estimate.  This is probably a
consequence of a difference in the prediction of the onset of
deformation in the $^{244}$Tl region.  (5) Fit to abundance pattern: The
scaling factor $f$ for the [U*f/U0] and [Th*f/Th0] age estimates has an error
$\Delta_{\rm logfac}$, originating from the discrepancies between the
calculated and the observed abundance patterns for stable r-process elements
(see Fig.~\ref{FigAb}).

\section{Results}

Fig.~\ref{FigAb} shows our calculated r-process elemental abundances.  In
Tab.~\ref{TabResults} we list our resulting age estimates together with the
various sources of uncertainties.

\begin{table}
\begin{center}
\caption{Age estimates and errors. Log($\epsilon_0$) is the predicted abundance
ratio produced in the r-process. $\Delta$'s refer to errors in the predicted
abundance ratios (see text).
\label{TabResults} }


\begin{tabular}{llllllllr}
\tableline
Ratio & Model & Log($\epsilon_0$) & $\Delta_\beta$ & $\Delta_{\rm par}$ &
$\Delta_{\rm exp}$
   & $\Delta_{\rm logfak}$ & $\Delta_{\rm total}$ & Age (Gy) \\
\tableline
U/Th       & ETFSIQ & -0.16 & 0.07 & 0.04 & 0.16 &     & 0.18 & 12.6$\pm$3.9 \\
U/Th       & HFBCS1 & -0.37 & 0.08 & 0.04 & 0.16 &     & 0.18 & 8.2$\pm$4.0 \\
U*f/U0   & ETFSIQ & -0.95 & 0.09 & 0.04 & 0.14 & 0.1 & 0.20 & 10.5$\pm$2.9 \\
U*f/U0   & HFBCS1 & -1.12 & 0.09 & 0.04 & 0.14 & 0.1 & 0.20 & 7.9$\pm$2.9 \\
Th*f/Th0 & ETFSIQ & -0.79 & 0.09 & 0.04 & 0.08 & 0.1 & 0.16 & 5.9$\pm$7.5 \\
Th*f/Th0 & HFBCS1 & -0.76 & 0.08 & 0.04 & 0.08 & 0.1 & 0.16 & 7.2$\pm$7.3 \\
U/Gd     & ETFSIQ & -0.78 & 0.10 & 0.04 & 0.23 &     & 0.25 & 10.5$\pm$3.7 \\
U/Gd     & HFBCS1 & -0.90 & 0.11 & 0.04 & 0.23 &     & 0.26 & 8.8$\pm$3.8 \\
Th/Gd    & ETFSIQ & -0.62 & 0.10 & 0.04 & 0.20 &     & 0.22 & 6.1$\pm$10 \\
Th/Gd    & HFBCS1 & -0.53 & 0.11 & 0.04 & 0.20 &     & 0.23 & 10.2$\pm$11 \\
Th/Ir    & ETFSIQ & -0.96 & 0.11 & 0.04 & 0.20 &     & 0.23 & 12.6$\pm$11 \\
Th/Ir    & HFBCS1 & -0.94 & 0.11 & 0.04 & 0.20 &     & 0.23 & 13.7$\pm$11 \\
\tableline
\tableline
\end{tabular}
\end{center}
\end{table}

The most robustly predicted abundance ratio is [U/Th], which represents
therefore the most reliable chronometer. Yet, the mass-model dependence is
still substantial, yielding ages of 13$\pm$4 Gyr (ETFSI-Q) or 8$\pm$4 Gyr
(HFBCS-1).  
$\beta$ delayed fission leads to significant corrections when
adopting the new Mamdouh et al. fission barriers
(+0.9 Gyr for the [U/Th] age, --0.8 Gyr for the [U/X] ages and 4 Gyr
(!) for the [Th/X] ages). While more theoretical work is needed, this
indicates that $\beta$ delayed fission should not be neglected.
While some of the predicted [U/X] and [Th/X] ages
agree well with [U/Th], others show discrepancies. The [U*f/U0] and [Th*f/Th0]
ages average over these discrepancies, and at least [U*f/U0] provides a
reasonable age estimate as well ([Th/X] suffer from large observational
uncertainties in the present data). However, we are now in the position to pick
the [U/X] and [Th/X] estimates that agree best with the [U/Th] age. These are
the ratios based on Gd and Ir abundances (Hill et al., this volume).  Our
predicted ratios for [Th/Gd] and [Th/Ir] can therefore be used for improved age
estimates of stars where no Uranium can be detected. In principle, other
observed [Th/X] ratios could be used together with our calculations, if an
appropriate correction factor would be applied.  Such a correction factor
can now be determined by requiring consistency with [Th/Ir] and [Th/Gd].
However, it will be crucial to verify whether the consistency of the age
estimates based on [U/Th], [U/Gd], and [U/Ir], as obtained from our r-process
model, holds for a larger sample of EMP stars.
 



\acknowledgments

This work was carried out under NSF contracts PHY 0072636 (Joint Institute for
Nuclear Astrophysics) and PHY 9528844. J.C. was supported by NSF grant
AST-9986974.

\end{document}